\documentclass[aps,twocolumn,prl,showpacs,amsmath,amssymb,10pt]{revtex4-1}
\usepackage{graphicx}
\usepackage{scrextend}
\usepackage{manfnt,pifont}
\usepackage{hyperref}
\usepackage{url}

\usepackage{url}
\usepackage[outline]{contour}
\usepackage{textgreek}

\newcommand{\bolddelta}{\textdelta}

\usepackage{CJK}

\begin{document}

\begin{CJK*}{GB}{}
\CJKfamily{gbsn}

\title{Six-Point AdS Gluon Amplitudes from Flat Space and Factorization}

\author{Luis F. Alday$^{a}$}
\author{Vasco Gon\c{c}alves$^{b}$}
\author{Maria Nocchi$^{a}$}
\author{Xinan Zhou$^{c}$}
\affiliation{$^{a}$Mathematical Institute, University of Oxford, Andrew Wiles Building, Radcliffe Observatory Quarter, Woodstock Road, Oxford, OX2 6GG, U.K.,}
\affiliation{$^{b}$Centro de F\'isica do Porto e Departamento de F\'isica e Astronomia, Faculdade de Ci\^encias da Universidade do Porto, Rua do Campo Alegre 687, 4169-007 Porto, Portugal,}
\affiliation{$^{c}$Kavli Institute for Theoretical Sciences, University of Chinese Academy of Sciences, Beijing 100190, China.}

\begin{abstract}
\noindent We present a powerful new approach to compute tree-level higher-point holographic correlators. Our method only exploits the flat-space limit, where we point out a novel and important simplification, and factorization of amplitudes in AdS. In particular, it makes minimal use of supersymmetry, crucial in all previous bootstrap methods. We demonstrate our method by computing the six-point super gluon amplitude of super Yang-Mills in AdS$_5$.

\end{abstract}

\maketitle
\end{CJK*}

\section{Introduction}
\noindent

Over the past few decades, there has been tremendous progress in our understanding and ability to compute scattering amplitudes in flat space-time. The exceptional simplicity of the Parke-Taylor formula \cite{Parke:1986gb}, contrasted to the complexity of intermediate diagrammatic computations, pointed at remarkable structures. Since then we have discovered many extraordinary properties, ranging from the relation between gravity and gauge theory amplitudes \cite{Bern:2010ue} to the appearance of positive geometry structures \cite{Arkani-Hamed:2017tmz}, and our understanding of scattering amplitudes keeps growing. Along the way, we also learned many useful lessons. 
One such lesson is that gauge theory amplitudes serve as building blocks for gravity amplitudes while being much simpler. Another lesson is that to uncover the full structure behind scattering amplitudes the computation of higher points is fundamental. 

In contrast, progress in AdS has been much slower, even at tree level, and is mostly restricted to the four-point case \cite{Rastelli:2016nze,Rastelli:2017udc,Alday:2020lbp,Alday:2020dtb}\footnote{See \cite{Bissi:2022mrs} for a review.}. While specific higher-point functions are known \cite{Goncalves:2019znr,Alday:2022lkk,Goncalves:2023oyx}, we lack tools to compute higher-point functions in AdS more generally. Meanwhile, in \cite{Alday:2021odx} a framework to study holographic gluon amplitudes, through CFT methods, was introduced. Already with four points, gluon amplitudes in AdS are much simpler than graviton amplitudes. In this letter, we present a powerful new approach to compute tree-level gluon amplitudes in AdS, which is tailored for higher points and systematically goes beyond previous approaches. The method consists of two steps. First, in the flat-space limit the AdS amplitudes reduce to flat-space amplitudes for a specific choice of polarizations. With this choice, not studied in the flat-space literature, amplitudes simplify drastically. Furthermore, the limit works not only for the full amplitude but also for individual diagrams, fixing much of the answer in AdS. In the second step, we go into AdS, demanding the correct factorization of the AdS amplitude into lower-point amplitudes. It turns out that this fixes the amplitude completely! We will demonstrate our method by computing the tree-level six-point amplitude in AdS$_5$. Note that a distinguishing feature of our new approach is that we make minimal use of supersymmetry which has been both the key and the bottleneck of previous methods. Hence the new method can be applied more generally, as will be shown in \cite{longpaper}.

\section{Kinematics}
\begin{figure}[h]
\centering
\includegraphics[width=0.35\textwidth]{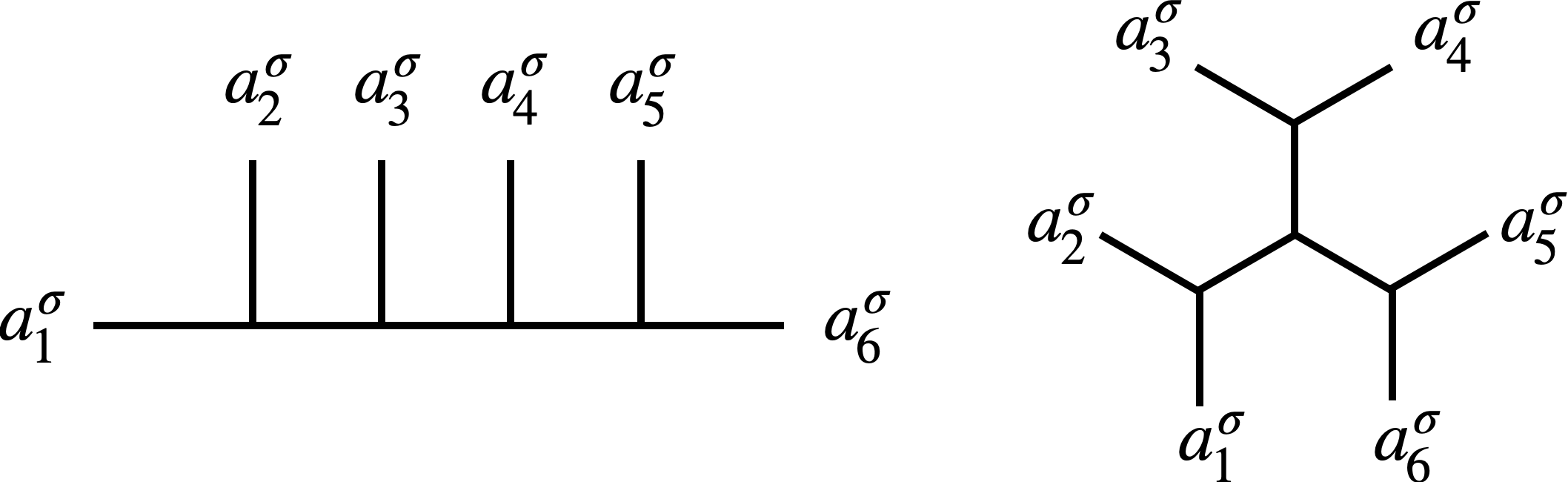}
\caption{Two topologies of cubic tree color diagrams, where $\sigma$ denotes a permutation of $\{a_1,a_2,a_3,a_4,a_5,a_6\}$.}
    \label{fig:color}
\end{figure}

We consider YM on AdS$_5$ with 4d $\mathcal{N}=2$ superconformal symmetry in the boundary parlance. While pure YM might be physically more relevant, the supersymmetric version is a more suitable testing ground for developing new technologies because the kinematics is more tractable. The theory contains a scalar field $s^a$ (super gluon) and a spin-1 gauge field $v^a_\mu$ as bosonic fields, transforming in the adjoint representation of a gauge group $G_F$ ($a=1,\ldots, {\rm dim}(G_F)$), as well as fermionic super partners. It can also be viewed as a consistent truncation of 8d $\mathcal{N}=1$ SYM on AdS$_5\times$S$^3$, which can be obtained by D3 branes probing an F-theory 7-brane singularity \cite{Fayyazuddin:1998fb,Aharony:1998xz} or D3 branes with D7 probes \cite{Karch:2002sh}. On the boundary, the super gluon is dual to a scalar field $\mathcal{O}^{a;\alpha_1\alpha_2}$, $\alpha_i=1,2$, with dimension $\Delta=2$ and transforms in the spin-1 representation of $SU(2)_R$. The gluon is dual to a conserved flavor current $\mathcal{J}^a_\mu$ which has $\Delta=3$ and is an R-symmetry singlet. From the boundary perspective, $G_F$ is a global flavor symmetry. Our target is to compute the six-point function of super gluons
\begin{equation}\label{defG6}
G_6(x_i;v_i)=\langle \mathcal{O}^{a_1}(x_1;v_1)\ldots \mathcal{O}^{a_6}(x_6;v_6)\rangle\;,
\end{equation}
where we have absorbed the $SU(2)_R$ indices by contracting them with R-symmetry polarization spinors
\begin{equation}
    \mathcal{O}^a(x;v)=\mathcal{O}^{a;\alpha_1\alpha_2}(x)v^{\beta_1}v^{\beta_2}\epsilon_{\alpha_1\beta_1}\epsilon_{\alpha_2\beta_2}\;.
\end{equation}
Compared to correlators with spinning operator insertions, the scalar correlator (\ref{defG6}) is kinematically simpler but captures essential features of AdS scattering. We will focus on tree level in AdS. Then  fermionic fields will not be exchanged and therefore are irrelevant. As on the LHS of (\ref{defG6}), we will often suppress the color indices to lighten the notation. But the color structures are described in the same way as in flat-space gluon amplitudes and are given by cubic tree color diagrams (Fig. \ref{fig:color}). Note that via the Jacobi identity, the snowflake diagrams 
\begin{equation}
    \mathbb{S}_{[a^\sigma_1a^\sigma_2a^\sigma_3a^\sigma_4a^\sigma_5a^\sigma_6]}=f^{a^\sigma_1a^\sigma_2b_1}f^{a^\sigma_3a^\sigma_4b_2}f^{a^\sigma_5 a^\sigma_6b_3}f^{b_1b_2b_3}\;,
\end{equation}
can be expressed in terms of the comb diagrams  
\begin{equation}
      \mathbb{T}_{[a^\sigma_1a^\sigma_2a^\sigma_3a^\sigma_4a^\sigma_5a^\sigma_6]}=f^{a^\sigma_1a^\sigma_2b_1}f^{b_1a^\sigma_3b_2}f^{b_2a^\sigma_4b_3}f^{b_3a^\sigma_5a^\sigma_6}\;.  
\end{equation}
In fact, $\mathbb{T}_{[a_1a^\sigma_2a^\sigma_3a^\sigma_4a^\sigma_5a_6]}$ with $\sigma$ being a permutation of $\{a_2,a_3,a_4,a_5\}$ form a basis \cite{DelDuca:1999rs}. While our perspective will be entirely from the bulk, we digress here to briefly comment on the boundary interpretation. The operators $\mathcal{O}^a$ are mesons in the dual gauge theory \footnote{E.g., the D3-D7 setup corresponds to 4d $\mathcal{N}=4$ SYM coupled to a small number of $\mathcal{N}=2$ fundamental hypermultiplets.}. The correlator (\ref{defG6}) at AdS tree level computes the leading $1/N$ contribution to the connected six-point meson correlator in a large $N$ gauge theory at infinite 't Hooft coupling.

\section{Mellin and factorization}
The best way to describe these holographic correlators is to use the Mellin space formalism \cite{Mack:2009mi,Penedones:2010ue}, which defines a scattering amplitude in AdS. In this formalism, we write 
\begin{equation}\label{defMellin}
 G_6=\int [d\delta_{ij}] \bigg(\prod_{i<j}x_{ij}^{-2\delta_{ij}} \Gamma[\delta_{ij}] \bigg)\mathcal{M}(\delta_{ij};v_i)\;,
\end{equation}
where $\mathcal{M}(\delta_{ij};v_i)$ is the Mellin amplitude and the Mellin variables satisfy the constraints $\delta_{ij}=\delta_{ji}$, $\delta_{ii}=-\Delta_i$, $\sum_{j\neq i}\delta_{ij} = 0$. It is most convenient to think of $\delta_{ij}$ as the Mandelstam variables formed from a set of fictitious flat-space momenta $\delta_{ij}=\bar{k}_i\cdot \bar{k}_j$ satisfying momentum conservation and on-shell condition $\bar{k}_i^2=-\Delta_i$. Then the constraints are automatically solved. Moreover, much of the flat-space intuition also extends to the Mellin amplitude. For example, Mellin amplitudes also enjoy factorization properties similar to flat-space amplitudes.  

\begin{figure}[h]
\centering
\includegraphics[width=0.45\textwidth]{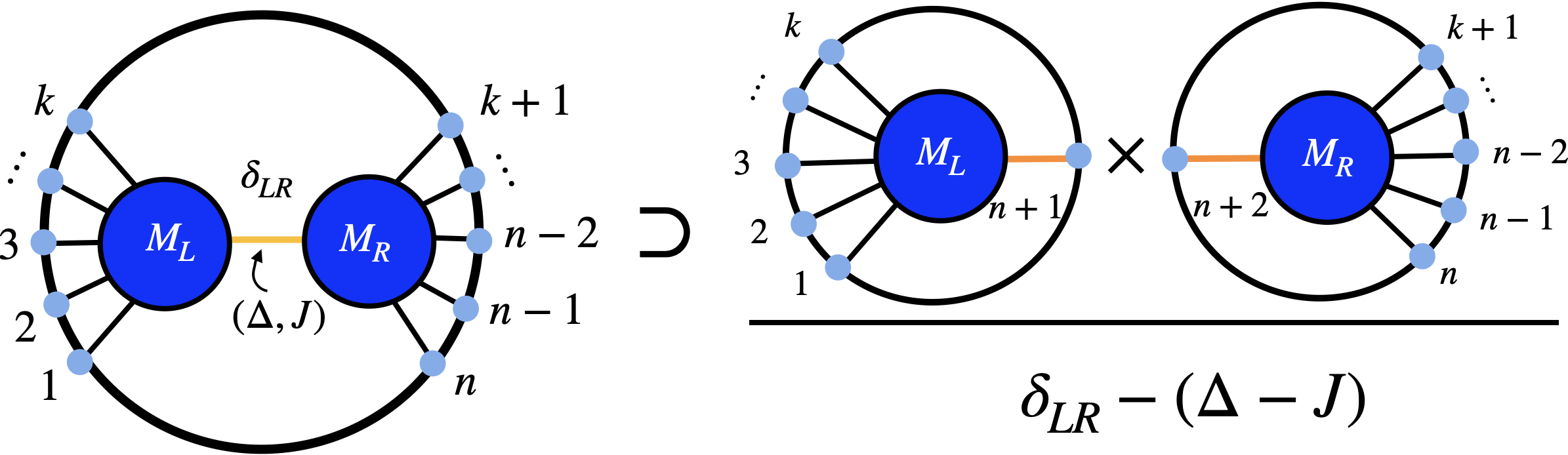}
\caption{The Mellin amplitude  factorizes into the product of two lower-point amplitudes at a pole.}
    \label{fig:factorization}
\end{figure}

More precisely, Mellin amplitudes are meromorphic functions with simple poles at $\delta_{LR}=\Delta-J+2m$, $m=0,1,2,\ldots$ that are associated with the exchange of an operator with dimension $\Delta$ and spin $J$. Here the propagator divides the $n$-point function into a $(k+1)$- and a $(n-k+1)$-point functions as in Fig. \ref{fig:factorization}, and $\delta_{LR}=\sum_{a=1}^k\sum_{i=k+1}^n\delta_{ai}$. The residues are controlled by lower-point amplitudes involving both the external and the exchanged operators \cite{Fitzpatrick:2011ia,Goncalves:2014rfa}. This is the CFT analogue of the well known factorization property of flat-space amplitudes. The simplest example is the exchange of a super gluon between two operators, say $\mathcal{O}_1$, $\mathcal{O}_2$, and the remaining ones. For this case the Mellin amplitude has a pole at $\delta_{12}=1$ given by  
\begin{align}
\mathcal{M}(\delta_{ij};v_i) \supset \frac{\mathcal{M}_{3}\mathcal{M}_{5}}{\delta_{12}-1}\;,
\end{align}
where $\mathcal{M}_{3}$, $\mathcal{M}_{5}$ are three- and five-point Mellin amplitudes of super gluons respectively.  Similar formulas also exist for the exchange of spinning operators, as well as for residues at satellite poles ($m>0$) \cite{Goncalves:2014rfa}.

\section{Simplification at flat space limit}
The Mellin amplitude contains information about flat space in the high energy limit. More precisely, the six-point scattering amplitude $\mathcal{A}^\flat$ of spin-1 gluons in flat space is related to the Mellin amplitude by \cite{Penedones:2010ue,Alday:2021odx}
\begin{equation}\label{MtoA}
  \lim_{\beta\to\infty} \beta \mathcal{M}(\beta s_{ij};v_i) \sim \mathcal{A}^\flat(s_{ij};e_i)\;,
\end{equation}
where $s_{ij}=k_i\cdot k_j$ are the flat-space Mandelstam variables. But coming from an AdS amplitude, the polarization vectors $e_i$ are restricted to a special  configuration where they are orthogonal to {\it all} momenta \cite{Alday:2021odx} 
\begin{equation}\label{epsilontransverse}
e_i\cdot k_j=0\;.
\end{equation} In fact, they lie within a four dimensional subspace and are related to the $SU(2)_R$ polarization spinors by 
\begin{equation}
    e^A_i=\frac{i}{\sqrt{2}}\sigma^A_{\alpha\beta}v^\alpha_i v^\beta_i\;,\quad A=0,1,2,3\;,
\end{equation}
where $\sigma^A_{\alpha\beta}$ are the Pauli matrices. This subspace is orthogonal to the subspace where $k_i$ live, which ensures the condition (\ref{epsilontransverse}). Note that the scalar super gluons and the spin-1 gluons in AdS lose their difference in the flat-space limit. They both become spin-1 gluons but with different polarizations. The flat-space amplitude $\mathcal{A}^\flat$ with polarizations obeying (\ref{epsilontransverse}) can be viewed as the dimensional reduction of the eight dimensional gluon amplitude into a scalar amplitude, in agreement with the picture of consistent truncation into AdS$_5$ SYM before taking the limit. 

\begin{figure}[h]
\centering
\includegraphics[width=0.35\textwidth]{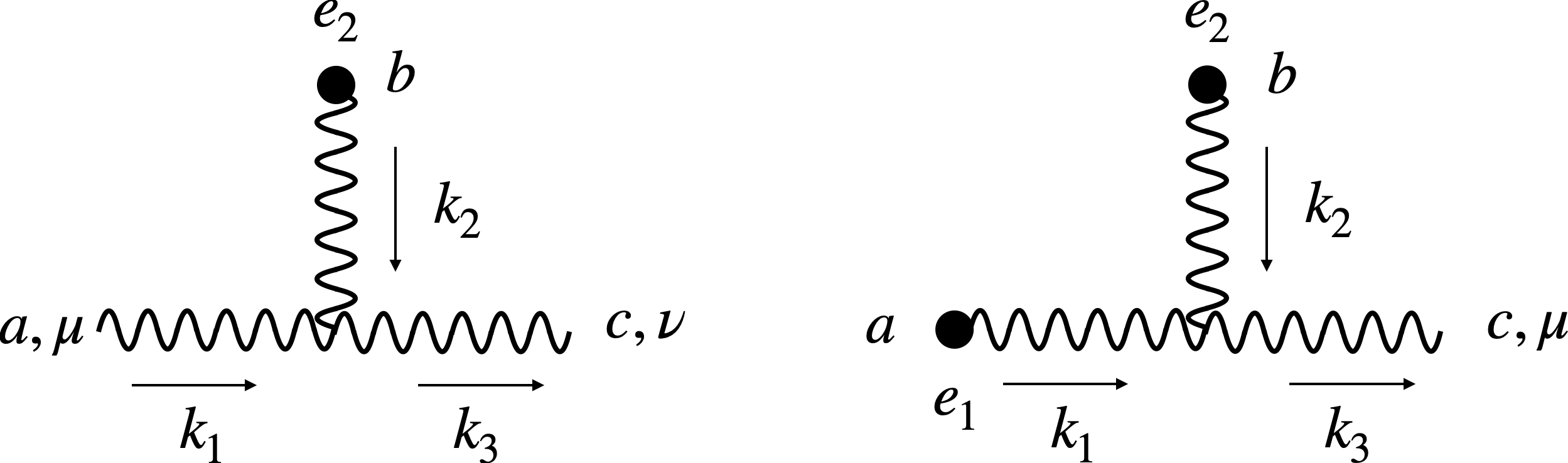}
\caption{Cubic vertices in flat space. The black dots are used to emphasize the contraction with the polarization vector $e$.}
\label{fig:flatcubicvertices}
\end{figure}

\begin{figure}[h]
\centering
\includegraphics[width=0.5\textwidth]{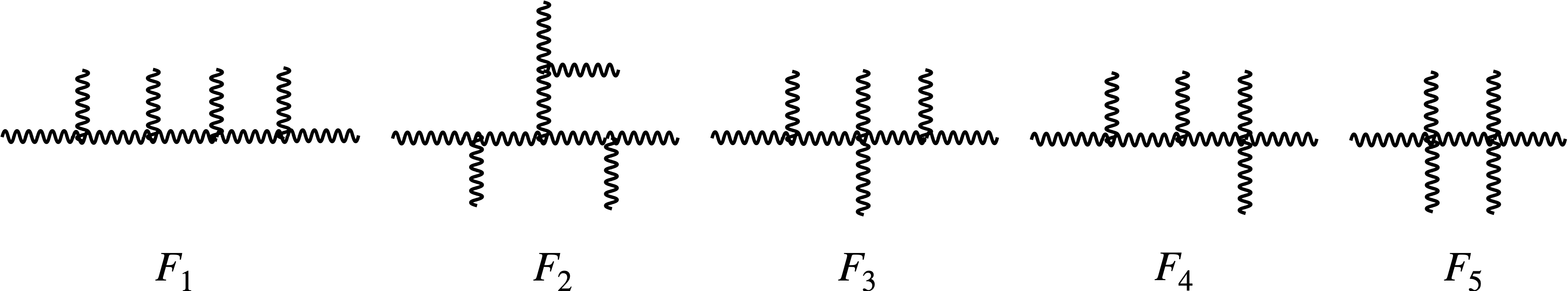}
\caption{All six-point Feynman diagrams in flat space.}
    \label{fig:FDs}
\end{figure}

The orthogonality constraint (\ref{epsilontransverse}) leads to significant simplifications to the flat-space amplitudes which  already can be seen at the level of Feynman rules. For example, the cubic vertices in Fig. \ref{fig:flatcubicvertices} are just $if^{abc}(-(k_2+k_3)_\mu e_{2,\nu}+(k_2-k_1)_\nu e_{2,\mu})$ and $if^{abc}(e_1\cdot e_2)(k_2-k_1)_\mu$. As a result, the contributing Feynman diagrams, listed in Fig. \ref{fig:FDs}, can all be easily computed and have a very simple form. For instance, the diagram $F_1$ (with $\sigma=1$ in comb diagram of Fig. \ref{fig:color}) is simply 
\begin{equation}\label{F1}  F_1=\mathbb{T}_{[a_1a_2a_3a_4a_5a_6]}\frac{e_{12}e_{34}e_{56}(s_{23}-s_{13})(s_{45}-s_{46})}{2s_{12}(s_{12}+s_{13}+s_{23})s_{56}}\;,
\end{equation}
where $e_{ij}=e_i\cdot e_j$. The other diagrams also have a similar level of complexity. This should be contrasted with Feynman diagrams with generic polarizations, where already for four points the expressions are quite cumbersome. The full flat-space amplitude is  
\begin{equation}
    \mathcal{A}^\flat = \sum_{i=1}^5 C_i F_i+{\rm permutations}\;,
\end{equation}
where $C_i$ are symmetry factors determined by combinatorics. However, one can also show that by imposing color-kinematic duality \cite{Bern:2008qj} on $\mathcal{A}^\flat$ all the $C_i$ coefficients are fixed up to an overall factor. The result fully agrees with the field theory limit of the six-point open string amplitude obtained from using pure spinor techniques \cite{Mafra:2010gj}\footnote{The explicit expression is available at \texttt{https://www.southampton.ac.uk/ $\sim $crm1n16/6pt-SYM\_bbbbbb.h}.}, by further imposing the orthogonality condition (\ref{epsilontransverse}). However, it is worth emphasizing again that the simplification from the configuration (\ref{epsilontransverse}) is drastic, allowing even for a diagrammatic evaluation by hand.

\section{Taxonomy of Witten diagrams}
\begin{figure}[h]
\centering
\includegraphics[width=0.5\textwidth]{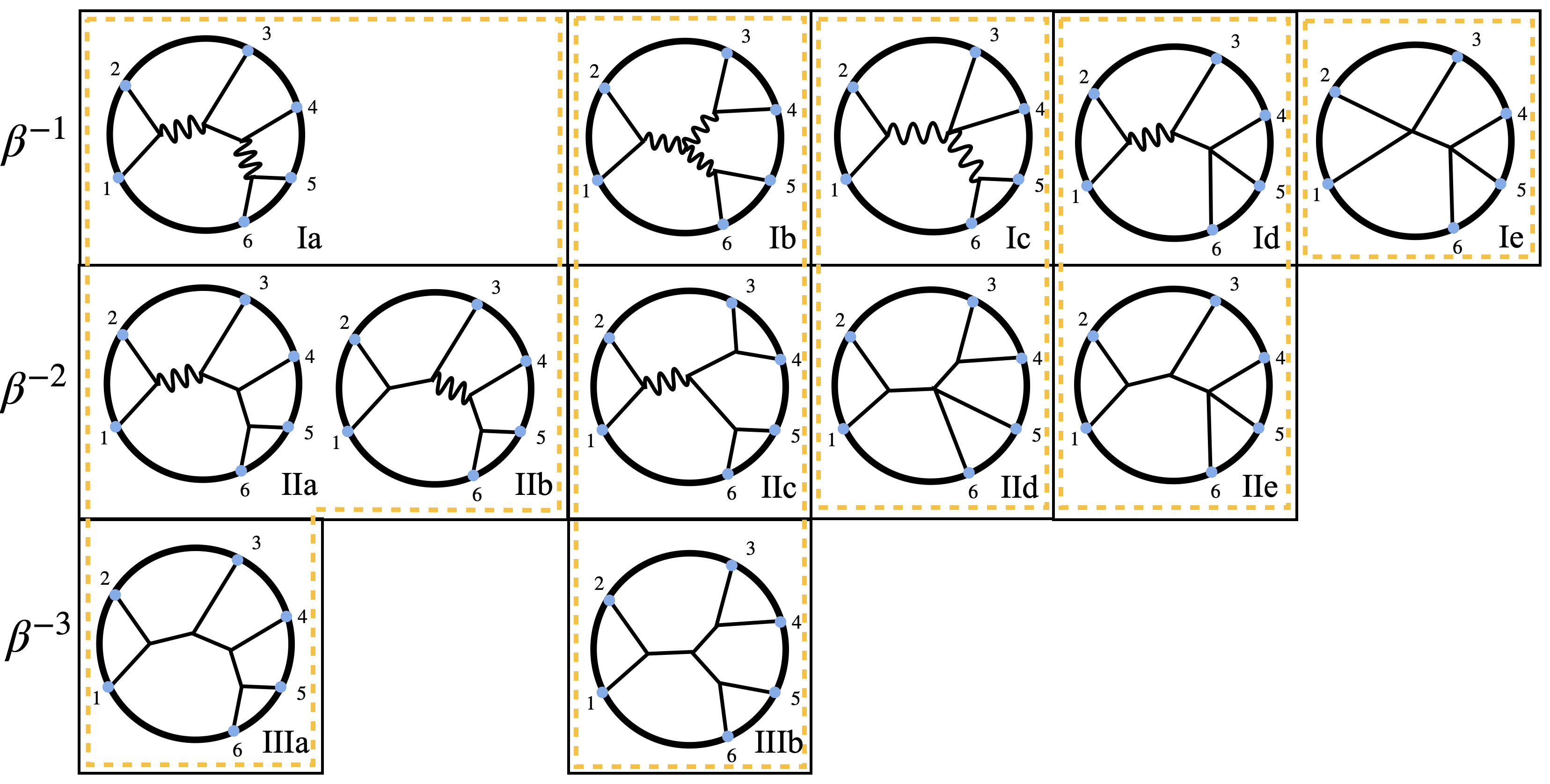}
\caption{Tree level Witten diagrams. Straight lines represent scalar super gluons ($s$) and wavy lines are spin-1 gluons ($v$). Diagrams in the $i$-th row decay as $\beta^{-i}$ in the large $\beta$ limit. Diagrams grouped together by the orange dashed line are related by replacing $v$ with $s$.}
    \label{fig:WDs}
\end{figure}

The AdS amplitude essentially is a collection of Witten diagrams. As we mentioned in (\ref{MtoA}), the Mellin amplitude reduces to the flat-space amplitude in the high energy limit. But this mapping is in fact more refined and holds at the level of individual diagrams. It is not difficult to check that the five Feynman diagrams in Fig. \ref{fig:FDs} correspond to the Witten diagrams in the first row of Fig. \ref{fig:WDs} respectively (see, {\it e.g.}, \cite{Penedones:2010ue,Fitzpatrick:2011ia,Bissi:2022mrs} for computation of Witten diagrams in Mellin space). For example, the Mellin amplitude of diagram Ia is 
\begin{equation}\nonumber
        \mathcal{M}_{\rm Ia}=\sum_{m=0}^1\frac{\mathbb{T}_{[a_1a_2a_3a_4a_5a_6]} e_{12}e_{34}e_{56}(\delta_{13}-\delta_{23})(\delta_{45}-\delta_{46})}{(\delta_{12}-1)(\delta_{12}+\delta_{13}+\delta_{23}+m-2)(\delta_{56}-1)}\;,
\end{equation}
which reduces to (\ref{F1}) in the flat-space limit. The finite sum over $m$ can be understood in terms of the truncation properties observed in \cite{Goncalves:2023oyx}.

\begin{figure}[h]
\centering
\includegraphics[width=0.4\textwidth]{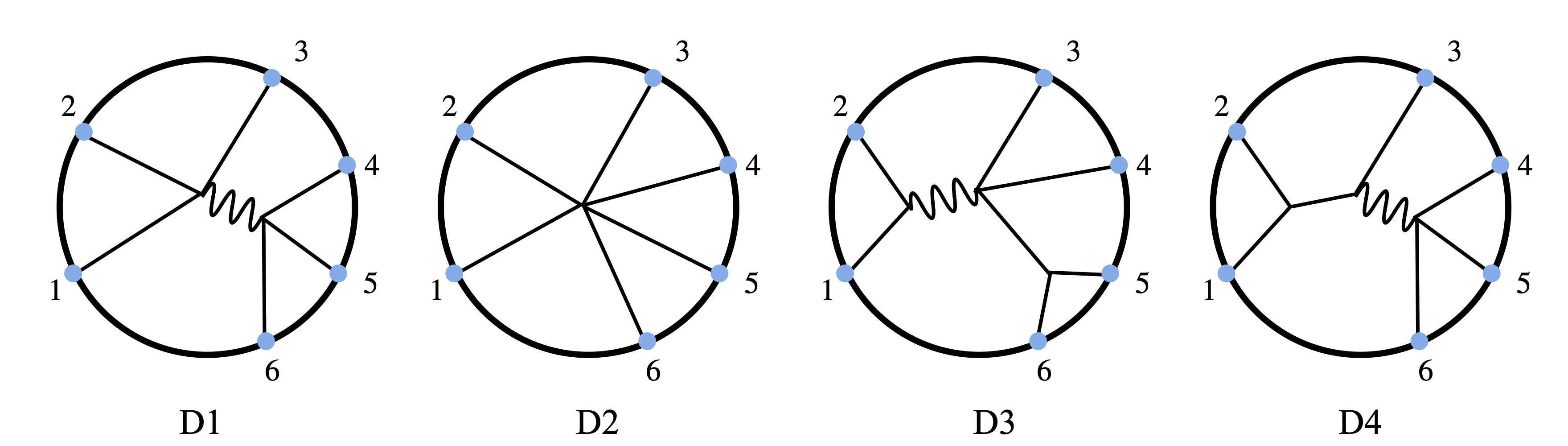}
\caption{Examples of  disallowed Witten diagrams.}
    \label{fig:disallowed}
\end{figure}

Note that the Mellin amplitudes of the Witten diagrams Ia to Ie
all scale as $\beta^{-1}$, as required by (\ref{MtoA}). Diagrams with a slower decaying behavior are disallowed as they violate the flat-space limit. For example, D1 and D2 in Fig. \ref{fig:disallowed} decay as $\beta^0$ and are therefore excluded. Note that the exclusion of D1 implies the nonexistence of quartic vertices with three scalars and one vector. This in turn implies that diagrams D3 and D4 do not exist and also decay as $\beta^{-1}$ \footnote{But already from matching the flat-space amplitude we can conclude that there are no such diagrams.}. But in the opposite direction, diagrams with faster decaying rates are not detected by the flat-space limit (\ref{MtoA}) and are not prohibited. These diagrams are cataloged in the second and third rows of Fig. \ref{fig:WDs} and are obtained from the first row by changing some vector internal lines into scalar lines. Note that some replacements violate R-symmetry ({\it e.g.}, replacing only one $v$ by $s$ in diagram Ib) and are therefore not listed. To summarize, Fig. \ref{fig:WDs} contains all possible six-point Witten diagrams up to permutations. 

\section{Bootstrapping six-point amplitude}
Using these ingredients, we can now formulate an efficient algorithm to compute the six-point Mellin amplitude. This comes in two steps. The flat-space limit $\mathcal{A}^\flat$ clearly allows us to fix the coefficient of each diagram in the first row of Fig. \ref{fig:FDs}. For the rest of the Witten diagrams, we observe that they have at least one internal scalar line which separates the diagram into a five-point diagram and a three-point diagram. Although these Witten diagrams are not captured by the flat-space limit, they are detected by the AdS amplitude factorization. We can therefore fix all  their contributions in terms of the five-point and three-point Mellin amplitudes of super gluons. To see how the strategy works in detail, it is instructive to first look at an explicit example where we reproduce the four-point function \cite{Alday:2021odx}. By different large $\beta$ scalings, we have
\begin{equation}
    \mathcal{M}^{\rm 4pt}_{\rm ansatz}=\mathcal{M}_{\rm L}+\mathcal{M}_{\rm SL}\;.
\end{equation}
The leading part consists of the gluon exchange diagrams and the contact diagrams
\begin{equation}\nonumber
\begin{split}
    \mathcal{M}_{\rm L}={}&\mathtt{c}_s \bigg[ \lambda_v\frac{e_{12}e_{34}(t-u)}{s-2}+\lambda_{c,1}e_{13}e_{24}+\lambda_{c,2}e_{14}e_{23}\bigg]\\
    +{}&(\text{t- and u-channels})\;,
    \end{split}
\end{equation}
where $\mathtt{c}_s=\mathbb{T}_{[a_1a_2a_3a_4]}$ and $\delta_{12}=\delta_{34}=\frac{4-s}{2}$, $\delta_{14}=\delta_{23}=\frac{4-t}{2}$, $\delta_{13}=\delta_{24}=\frac{4-u}{2}$. The relative coefficient between the exchange and contact diagrams is fixed by the flat-space limit to be $\lambda_{c,1}/\lambda_v=-1$, $\lambda_{c,2}/\lambda_v=1$. The subleading part is proportional to the scalar exchange
\begin{equation}\nonumber
    \mathcal{M}_{\rm SL}=\mathtt{c}_s \lambda_s\frac{e_{12}e_{34}-2v_{13}v_{24}v_{12}v_{34}}{s-2}+(\text{t- and u-channels})\;,
\end{equation}
where $v_{ij}=v_i^\alpha v_j^\beta \epsilon_{\alpha\beta}$. By factorizing the ansatz on the scalar internal line, we get a product of two super gluon three-point functions and this gives $\lambda_s/\lambda_v=2$.

The six-point function case is essentially the same, except for more complicated technical details which will be relegated to a separate publication \cite{Alday:2021odx}. Instead of starting from a sum of all the Witten diagrams in Fig. \ref{fig:WDs} with unfixed coefficients, it is more convenient to write an ansatz with the same analytic structure to avoid calculating the diagrams
\begin{align}\label{Mellinansatz}
    &\mathcal{M}_{\rm ansatz}=  \frac{P_{12;34;56}^{(2)}(\delta_{kl},v)}{(\delta_{12}-1)(\delta_{34}-1)  (\delta_{56}-1)}+\frac{P_{12;34}^{(1)}(\delta_{kl},v)}{(\delta_{12}-1)(\delta_{34}-1) }\nonumber\\
    &+ \frac{P_{12}^{(0)}(\delta_{kl},v)}{(\delta_{12}-1)}
+\sum_{m=0}^1\frac{C_{123,m}^{(2)}}{(\delta_{12}+\delta_{13}+\delta_{23})+m-2}+\textrm{perm}\;.
\end{align}
Here $P^{(n)}_{\dots}$ are polynomials of degree $n$ in the Mellin variables and $C^{(2)}_{\dots}$ are rational functions 
with the same analytic structure as the four-point Mellin amplitude.  The highest degree terms in the polynomials $P^{(n)}$ are easily fixed by matching with the flat-space scattering amplitude. In fact, the flat-space limit fixes completely $P_{12}^{(0)}$ and leaves $8$ undetermined coefficients inside the polynomials $P^{(2)}_{12;34;56}$ and $P_{12;34}^{(1)}$. These remaining coefficients, as well as those in $C^{(2)}_{123,m}$, are determined by the factorization of Mellin amplitudes \footnote{In addition to the factorization into $\mathcal{M}_5$ and $\mathcal{M}_3$, we can also consider the factorization of the amplitude on a scalar line into two four-point super gluon amplitudes. We used this to fix the relative coefficient of the contributions corresponding to the first row of Fig. \ref{fig:WDs} and the last two rows.}. 

\section{Consistency checks}
The six-point Mellin amplitude passes several nontrivial checks. The first check is that it factorizes correctly into a spinning five-point amplitude and a spinning three-point amplitude on an internal gluon line. Note that this condition was not part of the constraints imposed in our algorithm. More precisely, this factorization requires the six-point amplitude to contain a contribution 
\begin{equation}
\mathcal{M}(\delta_{ij};v_i) \supset \frac{\delta_{ai}\mathcal{M}_{3}^a\mathcal{M}_{5}^i}{\delta_{12}-1}\;,
\end{equation}
where the lower-point spinning amplitudes carry additional indices $a=1,2$, $i=3,4,5,6$, see the appendix. As we will discuss in more detail, the five-point spinning amplitudes can be fixed up to two undetermined coefficients by imposing only basic consistency conditions which include permutation symmetry, transversality, conservation, and factorization. The three-point amplitude is determined by these conditions up to an overall constant. The compatibility with our six-point amplitude provides a strong consistency check of our results.

Another independent check is the chiral algebra condition. In position space, superconformal symmetry imposes strong constraints on the form of holographic correlators, requiring them to satisfy certain meromorphy conditions when the R-symmetry polarizations are twisted \cite{Nirschl:2004pa,Beem:2013sza}. This condition is highly nontrivial and has been essential in previous approaches \cite{Rastelli:2016nze,Rastelli:2017udc,Alday:2020lbp,Alday:2020dtb,Goncalves:2019znr,Alday:2021odx,Alday:2022lkk}. At the same time, the chiral algebra condition is also practically very cumbersome to implement, especially at higher points. We will show that this property holds in \cite{longpaper}. But this condition is not needed in our algorithm, which is one of the reasons why the new algorithm is more powerful and efficient.

\section{Discussion}
In this letter we presented a new approach to compute higher-point holographic correlators. The algorithm relies on the flat-space limit and factorization, making only minimal use of superconformal symmetry. This greatly extends the range of holographic correlators which we can compute, including for instance SYM on AdS$_7\times$S$^3$, where a chiral algebra structure is not available. This method can also be used to compute the super graviton six-point amplitude in AdS$_5$. The six-graviton amplitude in the orthogonal configuration can be readily obtained by 
using the flat-space double copy relation \cite{Bern:2010ue}. Meanwhile, the super graviton five-point amplitude has also been computed in \cite{Goncalves:2019znr}. This provides the two necessary ingredients of our method for the super graviton six-point amplitude. It would also be interesting to further explore color-kinematic duality and double copy relation in Mellin space, extending previous observations at lower points \cite{Alday:2021odx,Zhou:2021gnu,Alday:2022lkk}. The appearance of subleading poles in (\ref{Mellinansatz}) suggests a generalized version of the duality in this case. 

The remarkable simple structure of the results in this letter make it plausible that a recursive method can be developed to compute higher-point correlators. On the one hand, the simplification of the flat-space amplitude in the orthogonal configuration certainly warrants more attention. It should be possible to obtain the all multiplicity result by adapting on-shell techniques. On the other hand, it can be checked that the five-point amplitude of super gluons \cite{Alday:2022lkk} can also be fixed from factorization if the four-point function with one spinning leg is used. Therefore, it would be important to systematically extend our analysis to spinning amplitudes after further developing the spinning Mellin formalism \cite{Goncalves:2014rfa}. In particular, this will allow us to apply the strategy to spinning gluon amplitudes in pure YM in AdS. Combining these elements, we can hope to generate higher-point AdS amplitudes directly from lower-point amplitudes.

\vspace{2cm}
We thank Carlos Mafra for helpful correspondence. The work of L.F.A. is supported by funding from the European Research Council (ERC) under the European Union's Horizon 2020 Research and Innovation Programme (grant agreement No 787185) and by the STFC grant ST/T000864/1. V.G. is supported by Simons Foundation grants \#488637 (Simons collaboration on the non-perturbative bootstrap) and Fundacao para a Ciencia e
Tecnologia (FCT) under the grant CEECIND/03356/2022. Centro
de F\'isica do Porto is partially funded by Funda\c{c}\~ao para a Ci\^encia e Tecnologia (FCT)
under the grant UID04650-FCUP. The work of M.N. is supported by funding from the Mathematical Institute, University of Oxford. The work of X.Z. is supported by the NSFC Grant No. 12275273,  the Fundamental Research Funds for the Central Universities, and starting funds from University of Chinese Academy of Sciences (UCAS), the Kavli Institute for Theoretical Sciences (KITS).

\section{Supplemental Material}
\subsection{Spinning Mellin amplitudes}

Spinning Mellin amplitudes are most convenient to define using the embedding space formalism where the action of the conformal group is linearized. Each point $x^\mu\in \mathbb{R}^d$ is lifted to a null ray $P^A\in \mathbb{R}^{1,d+1}$
\begin{equation}
    P\cdot P=0\;,\quad P\sim \lambda P\;.
\end{equation}
Using the rescaling symmetry, we can  gauge fix $P$ to be 
\begin{equation}
    P=\left(\frac{1+x^2}{2},\frac{1-x^2}{2},x^\mu\right)\;.
\end{equation}
Operators of dimension $\Delta$, spin $J$ are homogeneous functions of $P$ and $Z$
\begin{equation}
    \mathcal{O}(\lambda P,\alpha Z)=\lambda^{-\Delta}\alpha^J\mathcal{O}(P,Z)\;,
\end{equation}
where the polarization $Z^A\in \mathbb{R}^{1,d+1}$ encodes tensor structures and satisfies $P\cdot Z=0$. We further impose the transversality condition
\begin{equation}
    \mathcal{O}(P,Z+\beta P)=\mathcal{O}(P,Z)\;.
\end{equation}
For a correlator with $n$ scalars and 1 spinning (focusing on $J=1$) operator, the generalization of (\ref{defMellin}) is \cite{Goncalves:2014rfa}
\begin{equation}\label{defMspin}
\begin{split}
   {}& G_{n,1}(P_1,\ldots,P_n;P_0,Z)=\sum_{a=1}^n(Z\cdot P_a)\int [d\delta]\mathcal{M}_n^a \\
   {}& \times \prod_{i,j=1,i<j}^n \frac{\Gamma(\delta_{ij})}{(-2P_i\cdot P_j)^{\delta_{ij}}}\prod_{i=1}^n\frac{\Gamma(\delta_{0i}+\textrm{\bf{\bolddelta}}_{i}^{a})}{(-2P_i\cdot P_0)^{\delta_{0i}+\textrm{\bf{\bolddelta}}_{i}^{a} }}\;,
    \end{split}
\end{equation}
where $\textrm{\bf{\bolddelta}}_{i}^{a}$  is the Kronecker delta and the Mandelstam variables satisfy
\begin{equation}\nonumber
    \delta_{i0}=-\sum_{j=1}^n\delta_{ij}\;, \;\;\delta_{ij}=\delta_{ji}\;,\;\;\delta_{ii}=-\Delta_i\;,\;\;\sum_{i=1}^n\delta_{i0}=\Delta-1\;.
\end{equation}
The spinning Mellin amplitude is now a collection of partial amplitudes labelled by an index $a=1,\ldots,n$ because of the different structures $Z\cdot P_a$. When the current is conserved, we further have 
\begin{equation}
    \frac{\partial}{\partial P_0}\cdot\frac{\partial}{\partial Z}G_{n,1}(P_1,\ldots,P_n;P_0,Z)=0\;.
\end{equation}
In Mellin space, transversality and conservation translate to the constraints

\begin{equation}
   \begin{split}
   \sum_{a=1}^n\delta_{a0}\mathcal{M}_n^a={}&0\;,\\ \sum_{p,q=1,p\neq q }^n\delta_{pq}\big[\mathcal{M}_n^{a}\big]^{pq}={}&0\;, 
   \end{split}
\end{equation}
where 
\begin{equation}
    \big[f(\delta_{ij})\big]^{pq} = f(\delta_{ij}+\textrm{{\bf{\bolddelta}}}_{i}^{p}\textrm{{\bf{\bolddelta}}}_{j}^{q}+\textrm{{\bf{\bolddelta}}}_{i}^{q}\textrm{{\bf{\bolddelta}}}_{j}^{p})\;.
\end{equation}
To perform the first check mentioned in the main text, we need to consider the five-point spinning amplitude $\mathcal{M}_5^a$. The amplitude has poles at 
\begin{align}
\delta_{ij}=1\;,\, \ \  \delta_{0i}=0,1\;,
\end{align}
which correspond to the exchange of AdS fields in the OPE. Note that the appearance of two poles in $\delta_{0i}$ is a distinguishing feature compared to the scalar five-point amplitude where there is only one. However, this can be inferred from the structure of the OPE between the scalar and the current. Alternatively, the appearance of the additional pole can be expected from the consistency with the pole structure of the super gluon six-point amplitude. The five-point spinning Mellin amplitude is then just a rational function with finitely many poles. Moreover, this rational function should be consistent with the flat-space limit which in turn constrains the degree of the numerator.  We can make an ansatz where we also restore the R-symmetry polarizations and color structures which have been suppressed in (\ref{defMspin}). Note that the Mellin amplitude should have permutation symmetry among the four super gluon operators. Taking this into account, we arrive at an ansatz with 67 unfixed coefficients. Imposing transversality leaves 27 coefficients while  conservation further reduces the number to 9. By using the fact that the residues at the $\delta_{0i}$ poles are related to the known four-point spinning amplitude, we eventually have just two unfixed coefficients. Let us also comment that spinning four-point and three-point amplitudes can also be bootstrapped in the same way. The difference is that they are fully determined by permutation symmetry, transversality and conservation, up to an overall constant.

\bibliography{refg6pt} 
\bibliographystyle{utphys}
\end{document}